# Pressure-induced phase transitions in AgClO$_4$

D. Errandonea[1], L. Gracia[2], A. Beltrán[2], A. Vegas[3], and Y. Meng[4]

[1]Departamento de Física Aplicada-ICMUV, MALTA Consolider Team, Universitat de Valencia, Edificio de Investigación, c/ Dr. Moliner 50, 46100 Burjassot (Valencia), Spain

[2]Departament de Química Física I Analítica, MALTA Consolider Team, Universitat Jaume I, Campus de Riu Sec, 12080 Castelló, Spain

[3]Instituto de Química-Física Rocasolano, CSIC, c/ Serrano 119, 28006 Madrid, Spain

[4]HPCAT, Carnegie Institution of Washington, Advanced Photon Source, Building 434E, Argonne National Laboratory, 9700 South Cass Avenue, Argonne, IL 60439, USA

**Abstract:** AgClO$_4$ has been studied under compression by X-ray diffraction and density-functional theory calculations. Evidence of a structural phase transition from the tetragonal structure of AgClO$_4$ to a barite-type structure has been found at 5.1 GPa. The transition is supported by experiments and total-energy calculations. In addition, a second transition to a monoclinic structure is proposed to take place beyond 17 GPa. The equation of state of the different phases is reported as well as the calculated Raman-active phonons and their pressure evolution. Finally, we provide a description of all the structures of AgClO$_4$ and discuss their relationships. The structures are also compared with those of AgCl in order to explain the structural sequence determined for AgClO$_4$.

## I. Introduction

A number of structure types of $ABO_4$ oxides consist of $AO_8$ bisdisphenoids and $BO_4$ tetrahedra. They include important mineral structures such as zircon ($ZrSiO_4$), anhydrite ($CaSO_4$), silver perchlorate ($AgClO_4$), and scheelite ($CaWO_4$). These structures are closely and simply related via crystallographic operations [1] and, because of their compositional diversity and structural simplicity, played a key role in developing comparative crystal chemistry. After pioneering work in the 1960s – 1980s on the pressure effects on the structure of $ABO_4$ oxides [2, 3], there has been a large effort to determine systematic trends in pressure-induced phase transitions. Many zircon-structured oxides have been found to transform under compression either to a scheelite-type structure or to a monazite-type ($CePO_4$) structure [4–8]. In some cases, the zircon-monazite-scheelite sequence has been observed [9]. These transitions are consistent with crystal-chemistry arguments supporting that pressure increase can be mirror by changing the ratio of A/B cation sizes [10]. These arguments are also consistent with the anhydrite-monazite-barite ($BaSO_4$) high-pressure (HP) sequence found in $CaSO_4$ [11] and with several HP transitions detected in scheelites and related oxides [12 – 18]. In contrast with zircon, scheelite, and anhydrite, the effects of pressure on the crystal structure of silver perchlorate have not been studied yet. Here, in order to improve the understanding the HP behavior of $ABO_4$ oxides, we report a combined experimental and theoretical study of $AgClO_4$. Our results verify predictions of the HP structural systematic of $ABO_4$ oxides. In particular we observed one phase transition and identify the structure of the HP phase. In addition, the occurrence of a second transition is proposed and axial and bulk compressibility of different phases are determined.

## II. Experimental details

We performed HP X-ray diffraction experiments at room temperature up to 20.5 GPa. Studies were carried out with a symmetric diamond-anvil cell (DAC) with 480-µm-diameter culet diamonds and a 150-µm-drilled rhenium gasket, at 16-IDB beamline of the HPCAT at the Advanced Photon Source (APS). We used a monochromatic X-ray beam ($\lambda$ =0.36802 Å) focused down to 10 µm x 10 µm. The diffraction patterns were recorded on a MAR345 image plate located at 350 mm from the sample and integrated using FIT2D. To perform the experiments, we used micron-size fresh powder of anhydrous $AgClO_4$ with purity higher than 99.9% (Aldrich). The powder was loaded in the DAC using neon (Ne) as the pressure-transmitting medium to guarantee quasi-hydrostatic conditions in the pressure range covered by the experiments. Special care was taken during the loading to avoid hydration of $AgClO_4$. The pressure was measured by means of the ruby fluorescence technique. It was also confirmed with the equation of state (EOS) of Ne [19]. The indexation of the Bragg reflections was done with UNITCELL [20] and DICVOL [21]. GSAS [22] was used to carry out structural refinements.

## III. Overview of the calculations

Density-functional theory (DFT) calculations were performed with the CRYSTAL09 program package [23]. Chlorine and silver atoms have been described by HAYWLC-31G and HAYWSC-311d31G pseudo-potential basis set, respectively, while for oxygen atom it has been used the standard 6-31G*. The Becke's three-parameter hybrid non-local exchange functional [24] combined with the Lee-Yang-Parr gradient-corrected correlation functional, B3LYP [25], has been used. Hybrid density-functional methods have been extensively used for molecules and provide an accurate description of crystalline structures, bond lengths, and binding energies [26]. The diagonalization of the Fock matrix was performed at adequate *k*-points grids in the reciprocal space being the Pack-Monkhorst/Gilat shrinking factors IS = ISP = 4. The thresholds controlling the accuracy of Coulomb and exchange integrals were set to $10^{-8}$ and $10^{-14}$, whereas the percent of Fock/Kohn-Sham matrices mixing was set to 40 [23]. Fittings of the computed energy–volume data provide values of zero-pressure bulk modulus and its pressure derivative as well as enthalpy–pressure curves for the four studied polymorphs [27]. Vibrational-frequencies calculation in CRYSTAL is performed at the Γ point within the harmonic approximation, and the dynamical matrix is computed by numerical evaluation of the first-derivative of the analytical atomic gradients.

### IV.  Results

#### a.  X-ray diffraction experiments

At ambient conditions $AgClO_4$ has a tetragonal structure, space group (SG) $I\bar{4}2m$ (No. 121), with Z = 2 [28]. At high-temperature there is transition to a cubic structure, space group $F\bar{4}3m$ (No. 216), with Z = 4 [29]. X-ray diffraction measurements show that our samples have the tetragonal structure with unit-cell parameters *a* = 4.976(3) and *c* = 6.746(4) Å. These values and the obtained atomic coordinates (see Table I) agree with those previously reported [28]. The structure of $AgClO_4$ is shown in Fig.1. Following the classical description, it consists of layers of $ClO_4$ tetrahedra sharing corners with $AgO_8$ bisdisphenoids. The four Cl-O bond distances are 1.469 Å and the eight Ag-O bond distances are 2.528 Å (×4) and 2.726 Å (×4).

Fig. 2 shows a selection of diffraction patterns measured at different pressures. From ambient pressure up to 4.2 GPa all the diffraction patterns can be assigned to the low-pressure tetragonal structure. At 4.2 GPa three additional reflections appear as due to the solidification of Ne [29]. These peaks are identified with asterisks in the figure. From 4.2 GPa we increased the pressure up to 5.1 GPa. At this pressure we observed considerable changes in the diffraction pattern, indicating that in that pressure range a phase transition took place. No additional changes were detected up to the maximum pressure. We also found that the phase transition is non-reversible as can be seen in the upper trace of Fig. 2.

As described in the introduction, the HP structural sequence of many $ABO_4$ oxides can be understood considering crystal-chemistry arguments. According with these arguments, under pressure $AgClO_4$ (like $CaSO_4$) could take the structure of related sulphates and chromates that crystallize with the monazite- and barite-type structures. Considering these arguments, to analyze the HP data we considered as potential HP structures of $AgClO_4$: monoclinic monazite (SG *$P2_1/n$*), orthorhombic barite (SG *Pnma*), and $AgMnO_4$-type (SG *$P2_1/n$*), a monoclinic distortion of the barite-type structure. We found that the diffraction patters, collected at 5.1 GPa and higher pressure, could be indexed with an orthorhombic unit cell. Subsequent refinements of the diffraction pattern measured at 5.1 GPa showed that it could well match with a barite-type structure. The obtained structural parameters are given in Table II and are in good agreement with those of the aristotype barite [30] and also with those of the isostructural $KClO_4$ compound [31]. This result is consistent with the idea that oxides under compression prefer to take the structure of related compounds with larger cations [10]. According with our results the transition to the barite-type phase involves a volume collapse of 7 %, indicating that the transition is a first-order transformation. This fact is also consistent with the fact that the HP phase can be recovered at ambient pressure upon decompression. As can be seen in Fig. 1, the transition to the barite-type phase does not only involve an increase of the packing efficiency, but also an increase in the Ag coordination number from 8 to 8 + 4. In contrast the $ClO_4$ tetrahedra are only slightly modified. According with our experiments, at 5.1 GPa in the barite-type phase there are a double-degenerated short Cl-O distances (1.335 Å) and two longer distances 1.350 Å and 1.352 Å. In the other hand, for the Ag-O bonds we obtained the following distances: 2.537 Å (×2), 2.556 Å, 2.622 Å, 2.723 Å (×2), 2.753 Å (×2), 2.889 Å (×2), and 3.175 Å (×2). Therefore, the slightly distorted $ClO_4$ tetrahedra and the 12-coordinated Ag units resemble very much that of divalent-metal sulfate minerals isomorphic to barite (celestine, anglesite, and barite) [32].

From our experimental results we obtained the pressure evolution of the unit-cell parameters of the low- and high-pressure phase. The results are shown in Fig. 3 together with the pressure evolution of the volume. In this figure it can be seen that in the low-pressure phase the *c*-axis is more compressible than the *a*-axis. In the high-pressure phase the *a*-axis is also the most compressible one. The pressure–volume curves shown in Figure 3 were analyzed in the standard way using a third-order Birch–Murnaghan EOS [33]. The bulk modulus ($B_0$), its pressure derivative ($B_0´$), and the atomic volume ($V_0$) at zero pressure were obtained for both phases. For the low-pressure phase we got: $V_0$= 167(1) Å$^3$, $B_0$ = 29(4) GPa, and $B_0´$ = 4.1(2). For the high-pressure phase we got: $V_0$ = 310.9(9) Å$^3$, $B_0$ = 38(8) GPa, and $B_0´$ = 5.1(6). The obtained bulk moduli indicate that $AgClO_4$ is a quite compressible compound. The origin of this phenomenon is probably the weakness of the Ag-O bonds. Previously, it has been empirically determined that the bulk modulus in many $ABO_4$ compounds is related to the cation charge density of the $AO_8$ polyhedra [10]. In particular, it has been established that $B_0$ (GPa) = 610 Z/$d^3$, where Z is the cationic formal charge of A and *d* is the mean A–O distance at ambient pressure in Å. From our data we obtained an average Ag–O distance of 2.627 Å. Applying the formula given above, a bulk modulus of 33GPa is obtained, which agrees with the value obtained from the fitted EOS for the low-pressure phase of $AgClO_4$.

### b. Theoretical study of the structural properties

First-principle calculations support the experimental findings. Figs. 4 and 5 show the calculated total energy and enthalpy for relevant structures of $AgClO_4$. These are the tetragonal ambient-pressure structure ($I\bar{4}2m$), an orthorhombic barite-type structure (***Pnma***), and a monoclinic structure which belongs to space group ***P2$_1$/m*** (No. 11). This structure will be later described in detail. In addition to these three structures, based upon crystal-chemistry arguments, we also performed calculations for other candidate structures as scheelite, zircon, $AgMnO_4$-type, monazite, anhydrite, $KAlF_4$, ***Cmcm***-type $NaClO_4$, and HT $AgClO_4$. We found in agreement with calculations that the most stable structure at ambient conditions is the tetragonal structure of $AgClO_4$ ($I\bar{4}2m$). The calculated lattice parameters are summarized in Table III. They slightly overestimated the experimental values, although the differences are within the typical reported systematic errors in DFT calculations [34, 35]. Under compression, the orthorhombic barite-type structure becomes the most stable structure. According with the calculations, the tetragonal-to-barite transition takes place at 2 GPa. This pressure is slightly smaller than the experimental value (5.1 GPa). The difference can be caused by the presence of kinetic barriers [36]. The lattice parameters calculated for the barite phase at 5.4 GPa are summarized in Table III. Calculations underestimate the experimental value of the *a*-axis and overestimate the experimental value of the other two axes. As in the experiments, in the calculations we obtain a large volume collapse associated to the transition (6%), which indicates that the transition has a first-order character. This fact is consistent with the non-reversibility of the transition detected by experiments. From the calculations the pressure evolution of the lattice parameters and unit-cell volume, qualitatively agree with the experimental values as shown by the EOS parameters determined for the tetragonal (orthorhombic phase): $B_0 = 27$ GPa and $B_0´ = 4$ ($B_0 = 37$ GPa and $B_0´ = 5$).

Upon further compression, calculations predict a second transition to a monoclinic structure with symmetry *P*2$_1$/*m*. This structure was considered following the structural relationship between cations and alloys proposed by O'Keeffe and Hyde [37], later generalized by Vegas [38, 39]. Among the several possible structures, that of the HT phase of deuterated RbD$_2$PO$_4$ [40] was considered as one of the most plausible candidates, being so the object of the theoretical calculations, as described below.

According with these calculations, the second transition takes place at 17.2 GPa. Experiments were performed up to 20.5 GPa and this transition was not found. Probably it will be found upon further compression (note that the first transition was found in experiments at a pressure 3.2 GPa higher than calculated). Unfortunately, beyond 20.5 GPa the diffraction patterns broadens and distorts, not allowing to obtain any reliable results from them. Reasons for these changes are beyond the scope of this work and may be related either to the occurrence of pressure-induced phase transitions or to pressure-induced chemical decomposition of AgClO$_4$. The lattice parameters calculated for the HP monoclinic phase are summarized in Table III. Apparently the compression of this phase is isotropic. We found also that the monoclinic β angle is basically not affected by compression. According with calculations there is a volume change of about a 1.5 % at the orthorhombic-monoclinic transition. It also involves a change of the Ag coordination to 11, but the Cl coordination remains 4. For the monoclinic phase we determined the EOS being $B_0 = 43.4$ GPa and $B_0´ = 4$. Among all the structures included in the calculations the monoclinic phase remains the most stable one up to 40 GPa. In this structure, as in the other two, again it is observed that most of the compression of the crystal is accounted by the reduction of the Ag-O bonds, being the ClO$_4$ tetrahedra highly uncompressible units.

Changes in crystal structures induced by the phase transitions can be also related to different local atomic rearrangements in the crystals. Thus the higher density of the HP structures can be traced back to the unit-cell volume reduction due to a more effective packing of the O atoms surrounding the Ag atoms. To analyze pressure effects from this perspective we calculated the pressure-evolution of Ag-O and Cl-O bond distances for the three reported phases. Results are summarized in Figure 6. Calculated bond distances are slightly different than experimental distances. However calculations are capable to accurately describe the evolution and changes induced by pressure in atomic bonds. In the figure, clearly it can be seen that in the three phases the Ag-O bonds are more compressible than the Cl-O bonds; i.e. the ClO$_4$ tetrahedra behave like rigid units. The differential bond compressibility is more notorious in the low-pressure tetragonal phase. Another fact to remark is that at the first phase transition, the induced coordination change produces an enlargement of the Ag-O bonds in order to accommodate four oxygen atoms more surrounding the Ag atoms. Regarding the second transition, calculations show that the distortion of the ClO$_4$ tetrahedra is enhanced in the monoclinic phase and that the Ag-O bonds are strongly reduced at the orthorhombic-monoclinic transition. The average Ag-O bond distances are reduced a 5% at the transition.

There is a last comment we would like to add about theoretical results. As we already mentioned, calculations found the HT phase of AgClO$_4$ ($F\bar{4}3m$) to be not stable in comparison with the tetragonal phase. However, if we consider a volume expansion (negative pressures) we found that AgClO$_4$ prefers to take the cubic structure instead the tetragonal one. The HT $F\bar{4}3m$ phase [41] has a AgCl-subarray of the zincblende-type and is consistent with the stabilization of the cubic phase upon heating, as it was reported for many other compounds [15, 16]. This feature, in turn, agrees with the concept that relates oxidation and pressure [42] which will be discussed below. Following this concept, compression induces a volume and symmetry reduction in $AgClO_4$, but heating of the oxide releases pressure producing a reversion of the structural changes. Therefore, if we start the structural sequence in the HT structure of $AgClO_4$, volume reduction causes in $AgClO_4$ a gradual symmetry reduction following the sequence cubic-tetragonal-orthorhombic-monoclinic.

### c. Lattice-dynamics calculations

Lattice vibrations play an important role for materials modelling and their behaviour under pressure provides useful information regarding structural instabilities and phase transformations. The frequencies ($\omega$) of Raman-active modes for the tetragonal, orthorhombic, and monoclinic structures have been calculated as well as the Grüneisen parameters ($\gamma = B_0\, \partial \ln\omega/\partial P$) from its pressure dependences. Group theoretical considerations lead to 12 zone-center Raman-active modes in the tetragonal phase of $AgClO_4$: $\Gamma = 2A_1 + B_1 + 4B_2 + 5E$. The calculated ambient-pressure frequencies are summarized in Table IV. There it can be seen that the modes can be organized in three different groups. One group is composed by the six low-frequency modes with frequencies smaller than 251.5 cm$^{-1}$. These modes are separated by a phonon-gap of nearly 90 cm$^{-1}$ from three modes which follow the sequence $A_1 < E < B_2$. At high frequency, there are also three modes organized following the same symmetry order, which is separated from the second group by a phonon gap of about 150 cm$^{-1}$. The last three modes are associated to internal vibrations within the $ClO_4$ tetrahedra. In Table III we also reported the pressure coefficient ($\partial\omega/\partial P$) for each mode and the Grüneisen parameter. The two modes more affected by pressure are the $B_2$ mode with frequency 107.3 cm$^{-1}$ and the *E* mode with frequency 182.6 cm$^{-1}$. The last mode is associated with an Ag-O stretching. In addition, in the table it can be seen that the tetragonal structure presents two soft modes (at 64 and 251 cm$^{-1}$) characterized by a decrease of the vibrational frequency with pressure. These modes have symmetry *E* and *B$_1$* and both are associated to a bending between O-Ag-O units. Finally, we can see that Grüneisen parameters are much smaller in tetragonal AgClO$_4$ than in related compounds like zircon-type and scheelite-type ABO$_4$ oxides [43, 44]. In contrast they are comparable with those reported for barite-type ABO$_4$ oxides [45]. This indicates that in AgClO$_4$ and barite-type oxides the restoring force acting on an atom displaced from its equilibrium position is smaller than in other ABO$_4$ oxides. This is consistent with the fact that both barite-type oxides and AgClO$_4$ are the most compressible compounds within the ABO$_4$ family [46].

We will comment now on the Raman-active modes of the barite-type phase. Group theoretical considerations lead to 36 zone-center Raman-active modes: $\Gamma = 11A_g + 7B_{1g} + 11B_{2g} + 7B_{3g}$. The calculated sequence of modes resembles very much that of barite-type oxides. In our case, calculations show that lattice modes for frequencies smaller than 280 cm$^{-1}$ are basically ascribed to the motion of the Ag cation and tetrahedral ClO$_4$ units. The internal vibrations of the ClO$_4$ tetrahedron spanned the frequency range from 280 – 880 cm$^{-1}$, but with a gap between 406 and 609 cm$^{-1}$. The modes located between 610 and 785 cm$^{-1}$ can be associated to the $\nu_1$ and $\nu_3$ bands of barite [47]. For this structure the pressure coefficients of the different modes are comparable to those of the low-pressure phase. In particular, the modes with larger pressure coefficient are those located between 140 and 195 cm$^{-1}$. In addition, the barite structure presents one soft mode $A_g$ (at 375 cm$^{-1}$) characterized by a negative Grüneisen parameter which corresponds to a bending of O-Cl-O units (rotation of ClO$_4$ tetrahedra).

This feature is typical of scheelite-structure oxides [43], suggesting that at higher pressure orthorhombic phases should undergo a transition involving a strong coupling between a zone-centre optic mode and a strain of $A_g$ symmetry.

Finally, for the monoclinic phase, group theoretical considerations lead to 18 zone-center Raman-active modes: $\Gamma = 11A_g + 7B_g$. The calculated modes are summarized in Table IV. There it can be seen that modes more sensitive to pressure are those located between 200 and 300 cm$^{-1}$. In contrast with the other two structures there are no soft modes present in the monoclinic phase. Another typical feature of this phase is the increase of the frequency of the modes associated to internal vibrations of the ClO$_4$ tetrahedron in comparison with the other two phases. These frequencies increase following the structural sequence tetragonal-orthorhombic-monoclinic. This fact could be associated to the increase of the force constants of the internal vibrations.

V.  **Crystal chemistry of AgClO$_4$**

The two previously-known phases of AgClO$_4$ and the HP transitions can be rationalized when the structures are analyzed in terms of their cation sub-arrays (AgCl) and simultaneously compared with the structural changes undergone by the parent binary halide AgCl. This approach has proved to be fruitful in the analysis of many other structures of oxides [38, 39, 48] and is based on the following simple features:

i) There is a general trend by which the cation arrays in the oxides try to maintain the structure of the parent alloy/metal [38, 39].

ii) In many instances, the insertion of O atoms to form the oxides produces the same effect as the application of physical pressure, so that cations, in the oxides, can adopt structures typical of high-pressure phases of the alloy/metal. The behavior of the BaSn/BaSnO$_3$ pair of compounds is paradigmatic [42, 49].

iii) In those oxides that stabilize in structures associated to HP phases of the parent alloys, usually at high temperature (HT) a phase transition takes place, in such a way that the structure of the HT-phase recovers that of the ambient pressure phase of the alloy [39, 48].

iv) In many instances, the cation subarrays of oxides and their corresponding alloys have concurrent pathways in their phase transitions, as if the O atoms would not be present [48].

Figure 7 clearly shows that the tetragonal structure of $AgClO_4$ is a small distortion of the cubic rock-salt-type AgCl. The importance of this comparison resides in that the volumes of both unit cells is almost identical, with values of 170 and 167 Å$^3$ for $AgClO_4$ and AgCl, respectively. That means that the insertion of four O atoms per formula unit does not alter significantly the volume of the cell. The unique effect we might assign to the oxidation should be the tetragonal distortion of the rock-salt AgCl array. It should be remarked, however, that the distortion occurs in a concerted manner, so that the contraction of the $a$ and $b$ axes equals the elongation of the $c$ axis and hence, the average value of the three axes in the tetragonal $AgClO_4$ phase (5.567 Å) coincides with the unit-cell parameter of the halide AgCl (5.55 Å).

It has also been reported that, above 430 K, the tetragonal phase of $AgClO_4$ transforms into a cubic structure ($F\bar{4}3m$, $a$ = 6.95 Å, V =347 Å$^3$) where the cell volume is doubled and the O atoms show positional disorder [29]. In this structure the parent structure of the AgCl subarray is not modified resembling also rock-salt-type AgCl.

As discussed above, the tetragonal $AgClO_4$ converts at 5.1 GPa into a barite-type structure. Like $BaSO_4$, the structure is orthorhombic (*Pnma*) and can be described following O'Keeffe & Hyde [37] as an oxygen-stuffed AgCl-subarray of the FeB-type which is represented in Figure 8. The structure is formed by columns of trigonal prisms of Ag (Ba, Fe) atoms running parallel to *b* and where the ClO$_4$ (SO$_4$, B) groups/atoms are inserted. The final coordination number of the ClO$_4$ increases from 6 in the NaCl-like phase to 7 in the barite-type structure.

At this point, we can conclude that unlike the HT rock-salt-type and the tetragonal phases of AgClO$_4$, in which the parent structure of AgCl (NaCl-type) is maintained, the AgCl-subarray of the high pressure barite-type structure does not correspond to any known structure for the binary AgCl compound. This discrepancy, which at first glance could inquire the assertions quoted above, admits an explanation looking at the high-pressure behavior of AgCl [50, 51]. At increasing pressure AgCl undergoes the transition sequence NaCl → ***P*2$_1$/m** → TlI (*B*33) → CsCl (*a* = 3.20 Å). Although the *B*33 structure is also identified with CrB alloy, in this article, we will use the correspondence *B*33 ↔ TlI-type.

The reason for this choice is that, in the TlI-type structure, the Ag atoms (grey spheres in Figures 9d, f) are displaced from the center of the trigonal prisms towards the line connecting two Cl atoms. It is noteworthy that in TlI itself the Tl atoms are located just on that line, and that the same pattern is obtained whether the connected atoms are I or Tl. In this ideal case, both I and Tl atoms have a CN = 7. Looking at Figures 9d and 9f, we see that, in the isostructural AgCl-III the situation is almost identical. When the prisms are defined by Cl atoms (Figure 9) the Ag atoms lie into the prisms. On the contrary, when the prisms are formed by the Ag atoms, the Cl atoms deviate slightly (~ 0.2 Å) out the Ag-Ag line. It is a matter of choice to decide the best model. The factors which can influence these results are the applied pressure and the accuracy of the structure refinement depending, in turn, on the quality of the X-ray diffraction patterns.

In this context, two important features should be remarked. The first one is, that in some compounds, such as BaSn, the *B*33 → *B*2 transition is induced under pressure [44], so that the complete transition path becomes $B1$(NaCl) → $P2_1/m$ → TlI ($B33$) → $B2$ (CsCl) as it has been observed for AgCl itself [49, 50]. Recall that the double $B1$ → $B33$ → $B2$ transition has been observed in TlI itself and also in some alkali-metal halides. The second feature is that the FeB and the CrB (TlI)-type structures are intimately related. In fact, the FeB-type is considered as a stacking variant of the CrB-type structure. In fact, in the FeB boride itself, the $B33$-phase (*C*mcm) has been induced at low temperature [52].

The similarities between both structural types are illustrated in Figure 10. Note that the coordination polyhedron around the B atom, in FeB (*C*mcm) (Figure 10a) is quite similar to that surrounding the ClO$_4$ group in the barite-type AgClO$_4$ (FeB-type, *P*nma) represented in Figure 10b. On the other hand, when the barite structure is projected onto the (3 0 1) plane, one obtains the pattern drawn in Figure 10c which is part of the $B33$ structure represented in Figure 10d.

In this context, the structure of the high pressure phase AgCl-II, represented in Figures 9d and 11c, can be contemplated as an intermediate structure between the FeB and the $B33$ structures. Thus, if in the blue network of Figure 11c, the blocks of trigonal prisms would displace horizontally from each other, the $B33$ structure is formed. A similar result is obtained if the displacement occurs in the red network.

In Figure 9b we represented an alternative description of the HP phase AgCl-II ($P2_1$/m) which was not discussed above. This view of the structure is, in fact, a highly distorted $B1$ structure (Figure 9a). This fact is important because it gives sense to the complete transition path of AgCl under pressure i.e. $B1$ (NaCl) → $P2_1/m$ → TlI ($B33$) → $B2$ (CsCl), so that the $P2_1$/m phase can be seen as a $B1$ structure taking form to convert into the $B33$. As already shown, in AgCl–II ($P2_1$/m), fragments of both structures are recognizable.

To close this section we would like to make some comments on the predicted HP structure of the RbD$_2$PO$_4$-type. The theoretical calculations carried out on several structures concluded that a second HP-transition to be undergone by AgClO$_4$ should be the barite-type → RbD$_2$PO$_4$-type. This phase is monoclinic (***P*2$_1$/*m***) and its structure is represented in Figure 12a projected on the *ab* plane. In projection, the Ag atoms form rhombs which in 3D are highly distorted cubes (average edge *a* = 4.17 Å) centered by the ClO$_4$ groups. Consequently, the AgCl substructure is a distorted *B*2 (CsCl-type) structure which is represented in Figure 12b.

When the monoclinic structure of AgClO$_4$ is projected onto the *ac* plane, one can perceive that the distortion of the Ag cubes is such that preserves some structural features of a *B*33 structure (related to the FeB-type array of the barite structure) which appears as an intermediate phase in the process of convertion of *B1* into *B*2.

We have also analyzed the structure of other related compounds such as AgClO$_3$ which undergoes the transition ***P*4/*m*** → ***P*2$_1$3** at 413 K. In this cubic phase of AgClO$_3$, the AgCl subarray also forms a distorted *B*2 structure but with much more regular cubes. The structure is represented in Figure 12c and its average unit cell (*a* = 3.85 Å) is smaller than that of AgClO$_4$.

The magnitude of the distortions occurred in the two *B*2 structures of both AgClO$_4$ and AgClO$_3$ cannot be explained at present unless we admit the size of the respective anions as the dominant factor. What is important, however, is that the transitions sequence undergone by AgClO$_4$ can be identified with that of the parent AgCl, as summarized below:

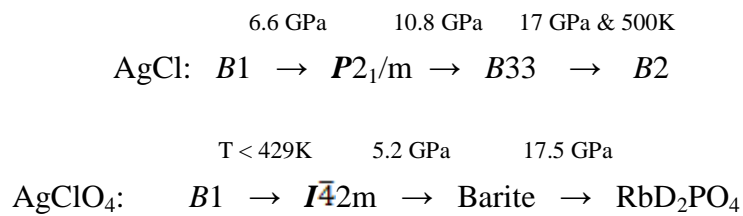

$$\text{AgCl:} \quad B1 \xrightarrow{6.6 \text{ GPa}} P2_1/m \xrightarrow{10.8 \text{ GPa}} B33 \xrightarrow{17 \text{ GPa \& 500K}} B2$$

$$\text{AgClO}_4: \quad B1 \xrightarrow{T < 429K} I\bar{4}2m \xrightarrow{5.2 \text{ GPa}} \text{Barite} \xrightarrow{17.5 \text{ GPa}} \text{RbD}_2\text{PO}_4$$

Note that in the above scheme the transitions of $AgClO_4$ start with the high temperature NaCl-type which at room temperature becomes distorted-NaCl. The lowering of temperature is equivalent to the application of pressure. These similarities provide new examples of how the structures of oxides are governed by cations and that these cations undergo their own phase transitions in spite of being embedded in an oxygen matrix.

## VI. Concluding Remarks

In this work we reported an experimental and theoretical study of the structural stability of $AgClO_4$ under compression. X-ray-diffraction experiments together with calculations have allowed us to determine the occurrence of a phase transition from the tetragonal structure of $AgClO_4$ to a barite-type structure. In addition, total-energy calculations predict the occurrence of a second transition to a structure with monoclinic symmetry. The whole sequence of structures reported for $AgClO_4$, indicates that their AgCl-subarrays follow a similar sequence as the HP phases of the parent AgCl. The room-temperature equations of state of different phases are also reported. Furthermore, lattice-dynamics calculations of Raman-active modes of different phases are also presented.

**Acknowledgements**

We acknowledge the financial support of the Spanish MICINN under Grants No. MAT2010-21270-C04-01 and No. CSD2007-00045. X-ray-diffraction experiments were performed at HPCAT (Sector 16), Advanced Photon Source (APS), Argonne National Laboratory. HPCAT is supported by CIW, CDAC, UNLV, and LLNL through funding from DOENNSA, DOE-BES, and NSF. A.P.S. was supported by DOE-BES, under Grant No. DE-AC02-06CH11357.

**Table I:** Unit-cell parameters and atomic coordinates for AgClO$_4$ obtained from powder diffraction at RT.

| $a = 4.976(3)$ Å and $c = 6.746(4)$ Å; $V = 167.0(3)$ Å$^3$; $Z = 2$ | | | | |
|---|---|---|---|---|
| Atom | Site | x | y | z |
| Ag | 2b | 0 | 0 | 0.5 |
| Cl | 2a | 0 | 0 | 0 |
| O | 8i | 0.1645(9) | 0.1645(9) | 0.1341(9) |

**Table II:** Unit-cell parameters and atomic coordinates for barite-type AgClO$_4$ obtained from experiments at 5.1 GPa.

| $a = 8.214(8)$ Å, $b = 5.160(5)$ Å, and $c = 6.651(6)$ Å; $V = 281.9(8)$ Å$^3$; $Z = 4$ | | | | |
|---|---|---|---|---|
| Atom | Site | x | y | z |
| Ag | 4c | 0.1822(2) | 0.25 | 0.3345(3) |
| Cl | 4c | 0.0697(1) | 0.25 | 0.8122(9) |
| O | 4c | 0.1902(2) | 0.25 | 0.9503(9) |
| O | 8d | 0.4033(4) | 0.5433(4) | 0.1961(2) |
| O | 4c | 0.4253(4) | 0.25 | 0.5901(5) |

**Table III:** Calculated parameters and atomic coordinates for different phases of AgClO$_4$. (top) low-pressure phase at ambient pressure, (center) barite phase at 5.4 GPa, (bottom) high-pressure monoclinic phase at 21.4 GPa.

| \multicolumn{5}{c}{$a = 5.091$ Å and $c = 6.944$ Å; V = 180.0 Å$^3$; Z = 2} | | | | |
|---|---|---|---|---|
| Atom | Site | x | y | z |
| Ag | 2b | 0 | 0 | 0.5 |
| Cl | 2a | 0 | 0 | 0 |
| O | 8i | 0.1829 | 0.1829 | 0.1416 |
| \multicolumn{5}{c}{$a = 7.934$ Å, $b = 5.473$, and $c = 7.139$ Å; V = 310.0 Å$^3$; Z = 4} | | | | |
| Atom | Site | x | y | z |
| Ag | 4c | 0.1823 | 0.25 | 0.2958 |
| Cl | 4c | 0.0660 | 0.25 | 0.8130 |
| O | 4c | 0.2351 | 0.25 | 0.9409 |
| O | 8d | 0.4295 | 0.5080 | 0.1768 |
| O | 4c | 0.3971 | 0.25 | 0.5597 |
| \multicolumn{5}{c}{$a = 6.201$ Å, $b = 4.953$ Å, and $c = 4.535$ Å, $\beta = 111.02°$; V = 130.0 Å$^3$; Z = 2} | | | | |
| Atom | Site | x | y | z |
| Ag | 2e | 0.2494 | 0.25 | 0.9453 |
| Cl | 2e | 0.2657 | 0.75 | 0.4617 |
| O | 2e | 0.4983 | 0.75 | 0.3881 |
| O | 2e | 0.3424 | 0.75 | 0.8503 |
| O | 4f | 0.1107 | 0.4899 | 0.3090 |

**Table IV:** Calculated phonon frequencies (cm$^{-1}$), pressure coefficients (cm$^{-1}$/GPa), and Grüneisen parameters for different phases of AgClO$_4$. Low-pressure phase at ambient pressure, barite phase at 5.4 GPa, and high-pressure monoclinic phase at 17.2 GPa.

| Tetragonal | | | | Orthorhombic | | | | Monoclinic | | | |
|---|---|---|---|---|---|---|---|---|---|---|---|
| Mode | ω | δω/δP | γ | Mode | ω | δω/δP | γ | Mode | ω | δω/δP | γ |
| $E$ | 11.5 | 0.28 | 0.66 | $B_{1g}$ | 11.0 | 0.43 | 1.56 | $B_g$ | 59.5 | 1.89 | 3.05 |
| $B_2$ | 35.4 | 0.35 | 0.27 | $A_g$ | 39.2 | 1.42 | 1.75 | $A_g$ | 85.6 | 2.55 | 2.65 |
| $E$ | 63.8 | -3.86 | -1.63 | $A_g$ | 60.8 | 3.61 | 4.02 | $B_g$ | 92.4 | 2.38 | 2.01 |
| $B_2$ | 107.3 | 5.65 | 1.42 | $B_{2g}$ | 33.1 | 0.85 | 0.86 | $A_g$ | 104.4 | 2.37 | 1.61 |
| $E$ | 182.6 | 7.97 | 1.18 | $B_{3g}$ | 52.2 | 0.33 | 0.24 | $A_g$ | 174.4 | 4.02 | 1.66 |
| $B_1$ | 251.5 | -1.07 | -0.11 | $B_{2g}$ | 60.7 | 1.85 | 1.26 | $B_g$ | 217.1 | 6.01 | 2.30 |
| $A_1$ | 339.1 | 3.26 | 0.26 | $B_{1g}$ | 84.2 | 1.62 | 0.69 | $B_g$ | 254.0 | 6.42 | 1.94 |
| $E$ | 357.0 | -0.02 | 0.00 | $A_g$ | 115.4 | 3.14 | 1.22 | $A_g$ | 238.3 | 5.34 | 1.58 |
| $B_2$ | 408.7 | 1.77 | 0.12 | $B_{2g}$ | 113.2 | 3.63 | 1.41 | $A_g$ | 289.6 | 5.87 | 1.35 |
| $A_1$ | 586.7 | 1.93 | 0.09 | $B_{1g}$ | 111.3 | 3.26 | 1.25 | $B_g$ | 348.7 | 3.69 | 0.56 |
| $E$ | 699.3 | 1.87 | 0.07 | $B_{3g}$ | 114.0 | 2.17 | 0.79 | $A_g$ | 335.9 | 1.75 | 0.25 |
| $B_2$ | 718.2 | 2.56 | 0.10 | $B_{2g}$ | 125.1 | 3.10 | 1.09 | $A_g$ | 391.1 | 0.97 | 0.11 |
| | | | | $B_{3g}$ | 140.0 | 4.21 | 1.42 | $B_g$ | 416.9 | 2.40 | 0.28 |
| | | | | $B_{1g}$ | 150.5 | 5.72 | 1.88 | $A_g$ | 429.2 | 0.81 | 0.09 |
| | | | | $A_g$ | 155.0 | 5.30 | 1.63 | $A_g$ | 584.5 | 0.13 | 0.01 |
| | | | | $B_{3g}$ | 168.7 | 6.00 | 1.70 | $A_g$ | 702.9 | 0.72 | 0.05 |
| | | | | $B_{3g}$ | 172.3 | 4.60 | 1.13 | $B_g$ | 801.1 | 4.10 | 0.24 |
| | | | | $A_g$ | 194.8 | 5.84 | 1.33 | $A_g$ | 811.8 | 1.87 | 0.10 |
| | | | | $B_{3g}$ | 283.8 | 1.36 | 0.18 | | | | |
| | | | | $B_{1g}$ | 286.1 | 0.99 | 0.13 | | | | |
| | | | | $A_g$ | 293.3 | 1.42 | 0.18 | | | | |
| | | | | $B_{2g}$ | 307.6 | 1.32 | 0.16 | | | | |
| | | | | $A_g$ | 369.5 | -0.41 | -0.04 | | | | |
| | | | | $B_{1g}$ | 375.8 | 0.23 | 0.02 | | | | |
| | | | | $B_{2g}$ | 382.3 | 0.49 | 0.05 | | | | |
| | | | | $B_{3g}$ | 386.6 | 0.86 | 0.08 | | | | |
| | | | | $A_g$ | 393.5 | 0.77 | 0.07 | | | | |
| | | | | $B_{3g}$ | 406.3 | 0.65 | 0.06 | | | | |
| | | | | $B_{3g}$ | 609.9 | 2.53 | 0.16 | | | | |
| | | | | $A_g$ | 614.5 | 2.84 | 0.18 | | | | |
| | | | | $B_{2g}$ | 710.1 | 4.37 | 0.24 | | | | |
| | | | | $B_{1g}$ | 718.0 | 4.42 | 0.24 | | | | |
| | | | | $A_g$ | 734.3 | 4.27 | 0.22 | | | | |
| | | | | $B_{3g}$ | 776.3 | 4.20 | 0.21 | | | | |
| | | | | $A_g$ | 758.8 | 0.36 | 0.02 | | | | |
| | | | | $B_{3g}$ | 784.6 | 1.92 | 0.09 | | | | |

**Figure captions**

**Figure 1:** (Color online). Different structures of AgClO$_4$: (a) ambient pressure tetragonal structure, (b) barite-type, and (c) HP monoclinic. Different polyhedra are shown. Ag: large grey spheres, Cl: medium green spheres, and O: small red spheres.

**Figure 2:** Selection of X-ray diffraction patterns. Pressures are indicated. (r) indicates the pattern collected after pressure release. Neon peaks are shown identified with asterisks. Background was subtracted.

**Figure 3:** Pressure evolution of unit-cell parameters and volume of AgClO$_4$. In the upper part the lines are a guide to the eye, in the lower part they represent the obtained EOS. For the barite-type phase we plotted V/2 to facilitate comparison.

**Figure 4:** (Color online) Total energy as a function of volume per formula unit for the tetragonal structure, barite-type and HP monoclinic.

**Figure 5:** (Color online) Enthalpy versus pressure curve for tetragonal, barite-type and HP monoclinic structures (taking the tetragonal structure as reference).

**Figure 6:** Pressure evolution of Ag-O and Cl-O distances for the different phases of AgClO$_4$.

**Figure 7:** (color online) (a) tetragonal structure ($I\bar{4}2m$) of AgClO$_4$ at ambient conditions ($a$ = 4.976 and $c$ = 6.746 Å). (b) rock-salt structure ($F\bar{4}3m$) of AgCl at ambient conditions ($a$ = 5.55 Å). Ag (grey), Cl (green), and O (red).

**Figure 8:** (color online) (a) barite-type structure of AgClO$_4$ (*Pnma*) viewed along *b*. The trigonal prism centering the cell in (a) has been magnified in (b) to outline the complete coordination polyhedron. Ag (grey), Cl (green), and O (red). The Ag-Ag contacts are drawn with blue lines. (c) The structure of the FeB alloy, projected on the same plane, to show its similarity with the cation array of AgClO$_4$. Fe-green, B-yellow.

**Figure 9**: (color online) Sequence of structures of AgCl by increasing pressure. (a) ambient pressure rock-salt AgCl structure. (b) high-pressure $P2_1/m$ phase also called AgCl-II. (c) $B33$ structure assigned to CrB- and TlI-type. (d) Alternative description of the AgCl-II structure viewed as a distortion of the $B33$ structure drawn in (b). (e) The same as in (d) but out of plane to show the trigonal prisms of Cl atoms centered by the Ag atoms. (f) $B33$ structure rotated 90º around the $b$ axis to show its similarities with the AgCl-II. (g) $B2$ (CsCl-type) structure of AgCl.

**Figure 10:** (color online) (a) Coordination polyhedron around the B atom in the $B33$ phase of FeB. (b) The same polyhedron formed by Ag atoms around the ClO$_4$ group in barite-type AgClO$_4$. (c) Fragment of the barite-type structure projected onto the (3 0 1) plane showing its similarity with the HP $B33$ structure of AgCl–III, represented in (d).

**Figure 11:** (color online) (a) The anti-structure of AgCl-III ($B33$), viewed along $a$. Ag atoms (grey) are connected with red lines to show the trigonal prisms occupied by Cl atoms (green). Compare with Fig. 9c. (b) The same structure, viewed along the $c$ axis, showing how that Cl atoms are slightly out of the trigonal prisms. Compare with Fig. 9f. (c) The structure of monoclinic AgCl-II, viewed along the $b$ axis. Both structure and anti-structure have been drawn with red and blue lines, respectively. Note that the red network is the same as that drawn in Fig. 9d. The blue lattice shows that the Cl atoms are slightly out of the horizontal trigonal prisms but inserted into the vertical trigonal prisms.

**Figure 12:** (color online) (a) HP structure of AgClO$_4$ predicted by calculations. It is monoclinic ($P2_1/m$) and isostructural to RbD$_2$PO$_4$. The Ag atoms (grey) form a distorted simple-cubic array centered by the ClO$_4$ groups [Cl (green) and O (red)]. (b) An isolated Ag$_8$ cube showing the distorted CsCl-type structure formed by the AgCl subarray. (c) The cubic HT-phase of AgClO$_3$ showing the CsCl-type structure. Colors as in (a).

**Figure 1**

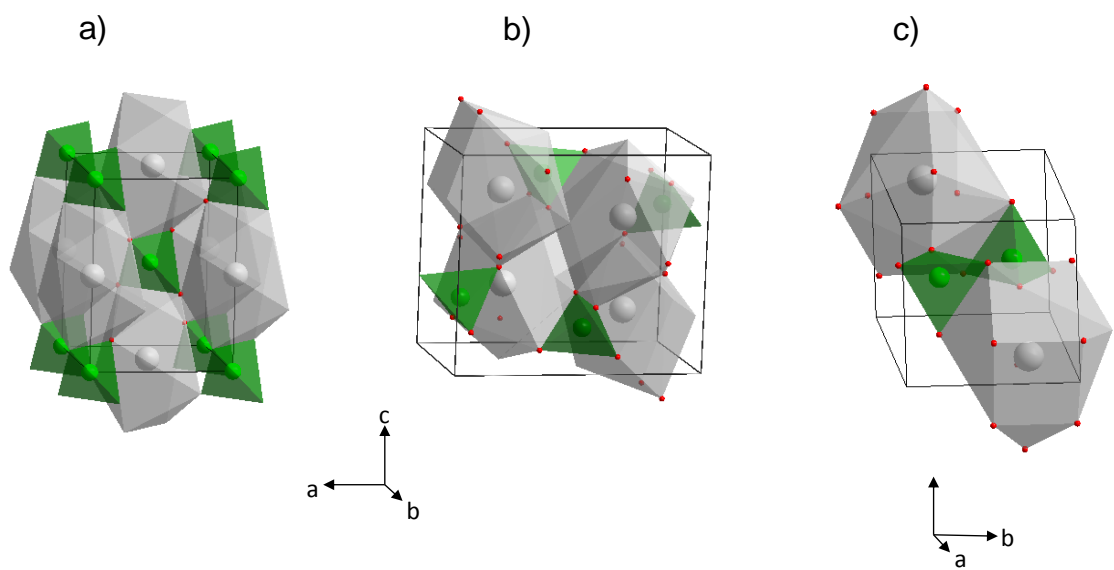

**Figure 2**

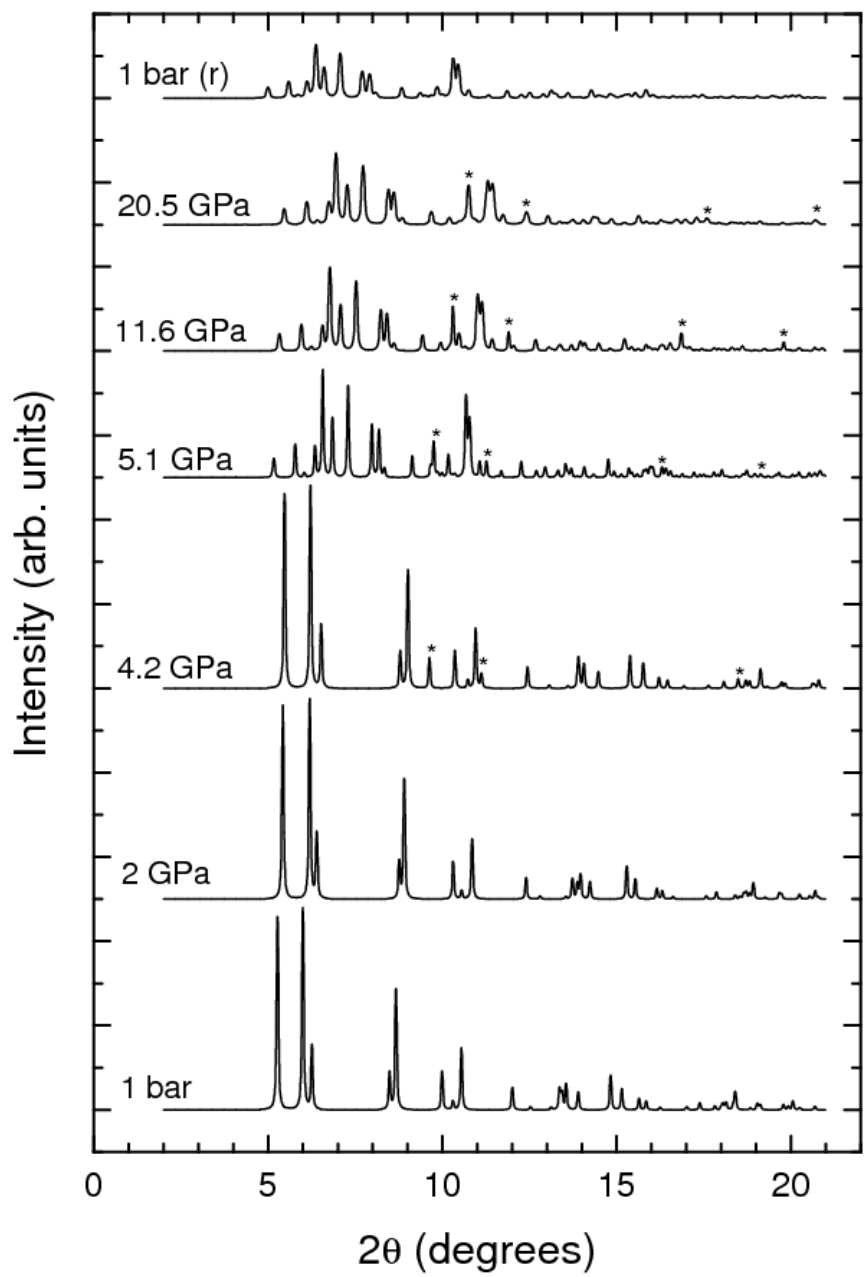

**Figure 3**

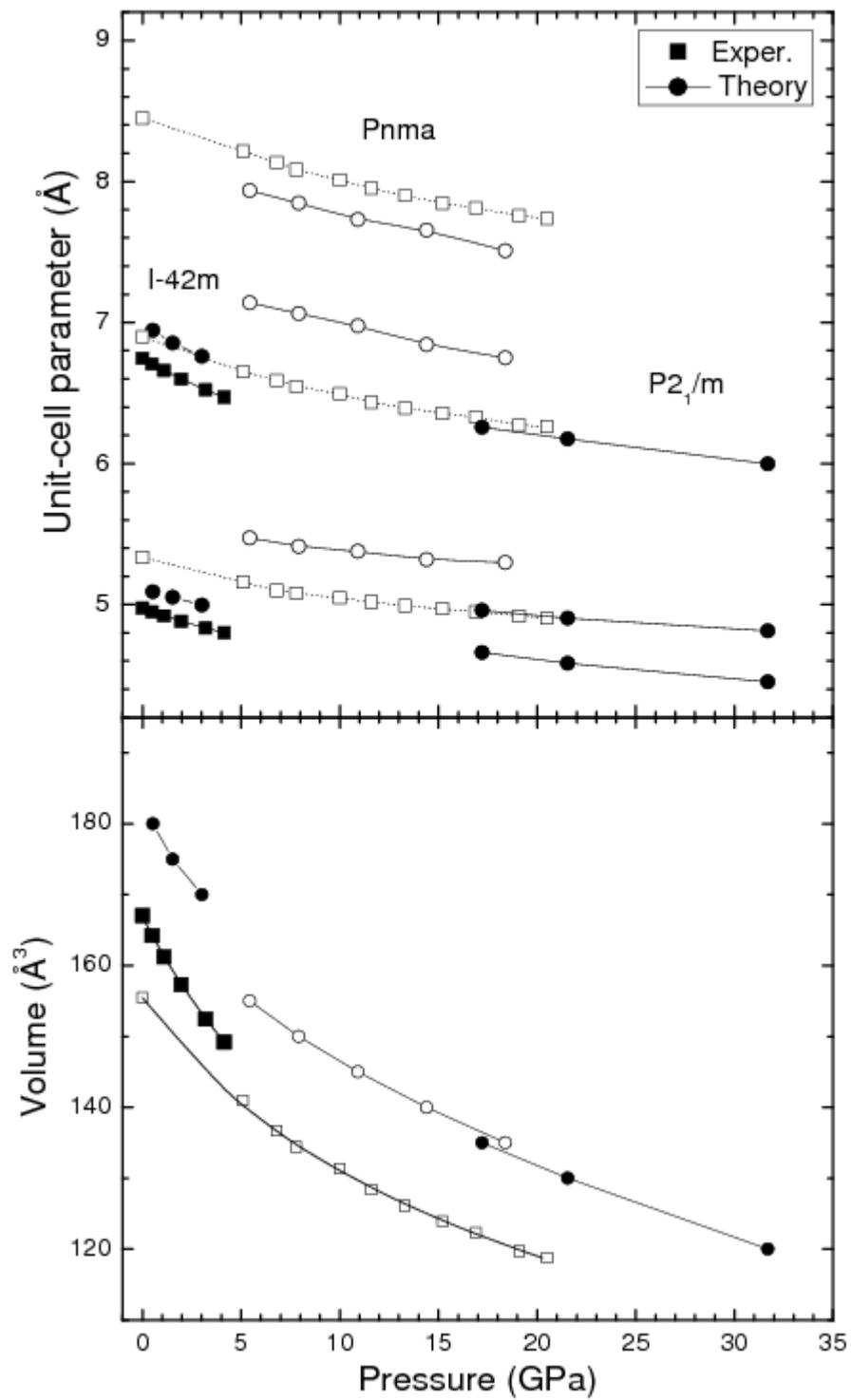

**Figure 4**

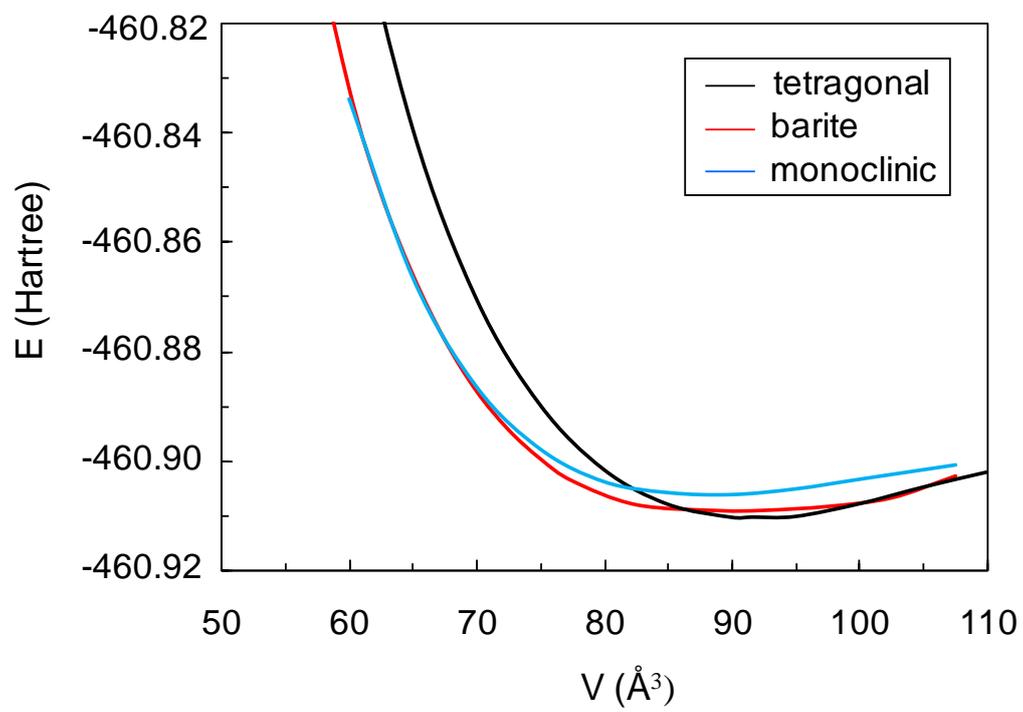

**Figure 5**

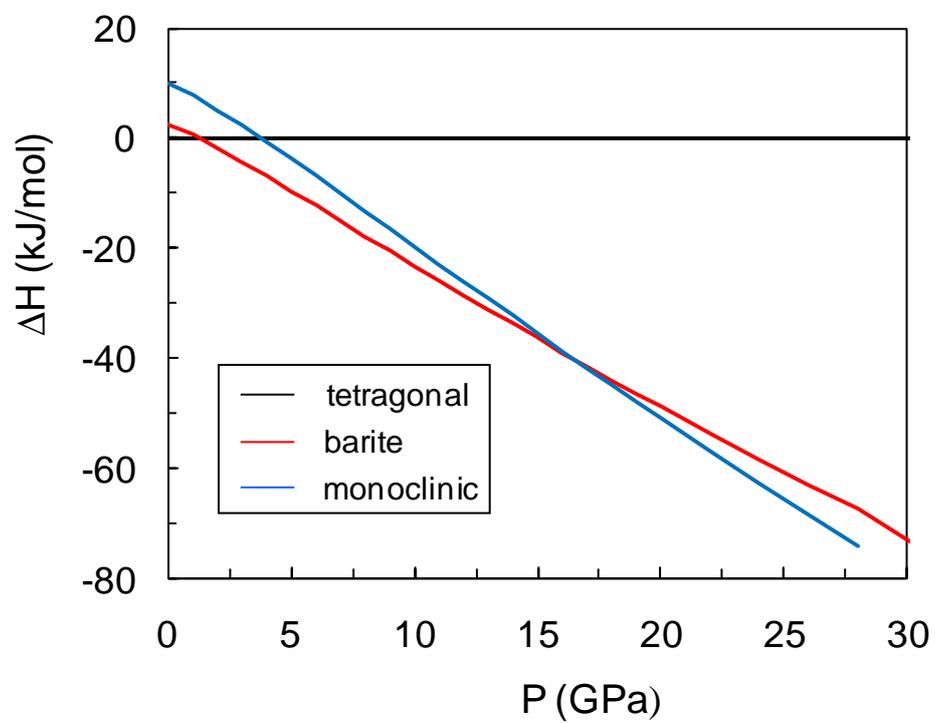

**Figure 6**

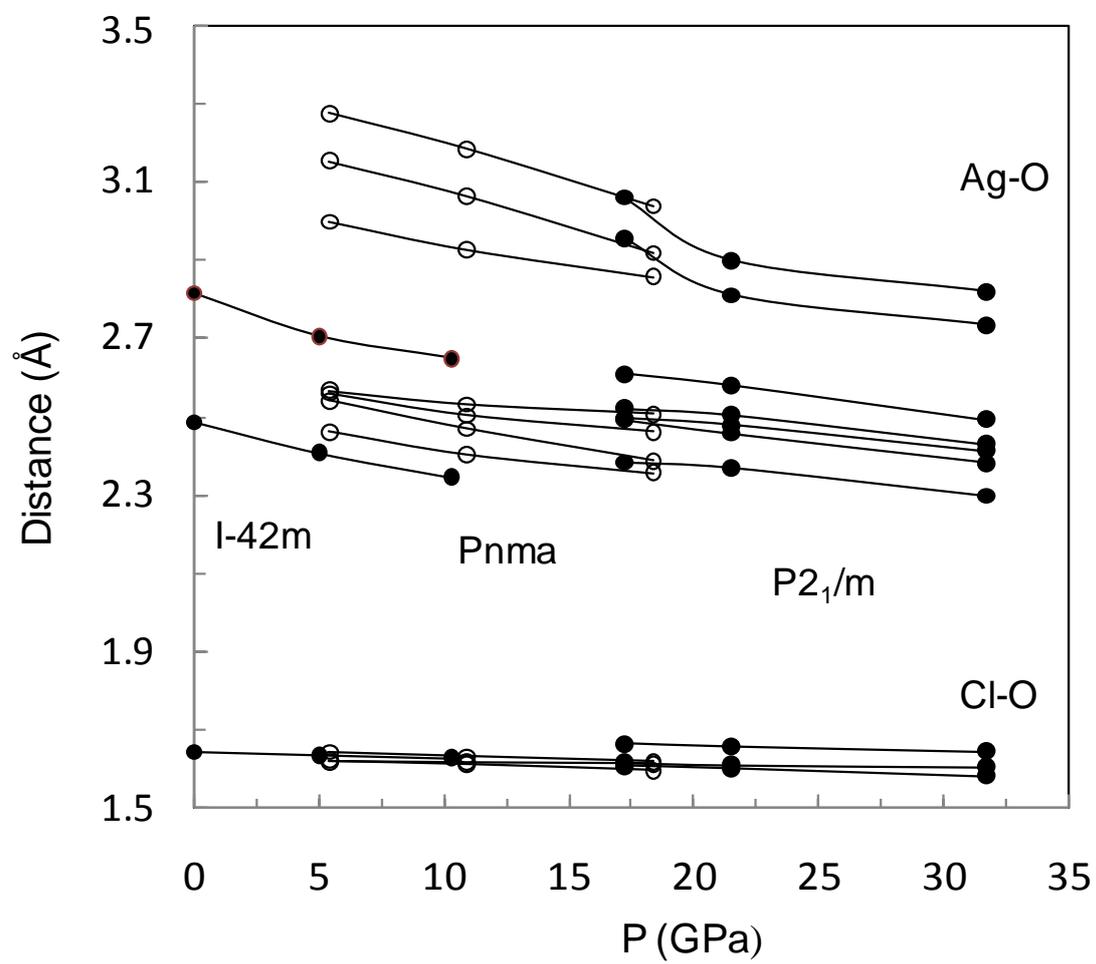

**Figure 7**

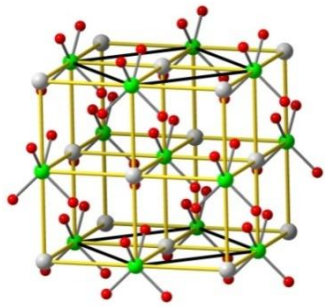  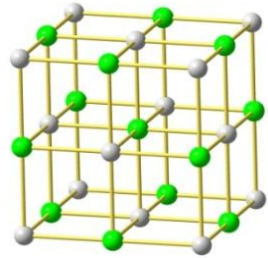

(a)　　　　　　　　　　　　(b)

**Figure 8**

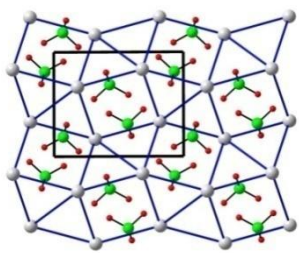 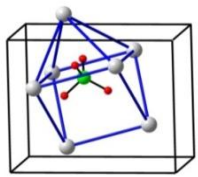 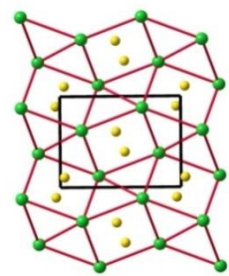

       (a)               (b)              (c)

**Figure 9**

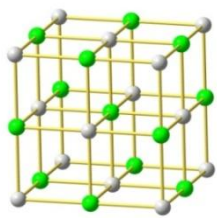 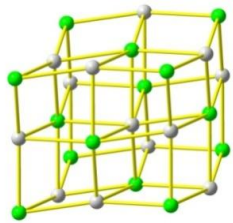 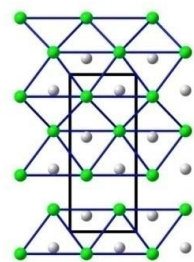

(a)        (b)        (c)

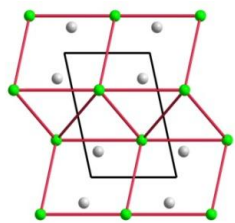 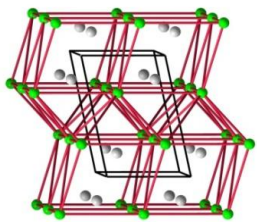 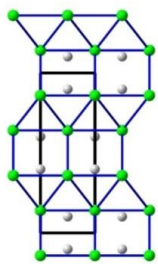 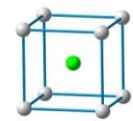

(d)        (e)        (f)        (g)

**Figure 10**

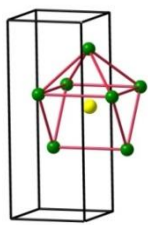 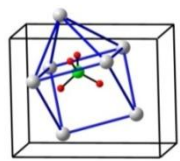 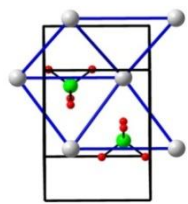 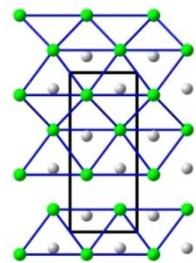

(a)  (b)  (c)  (d)

**Figure 11**

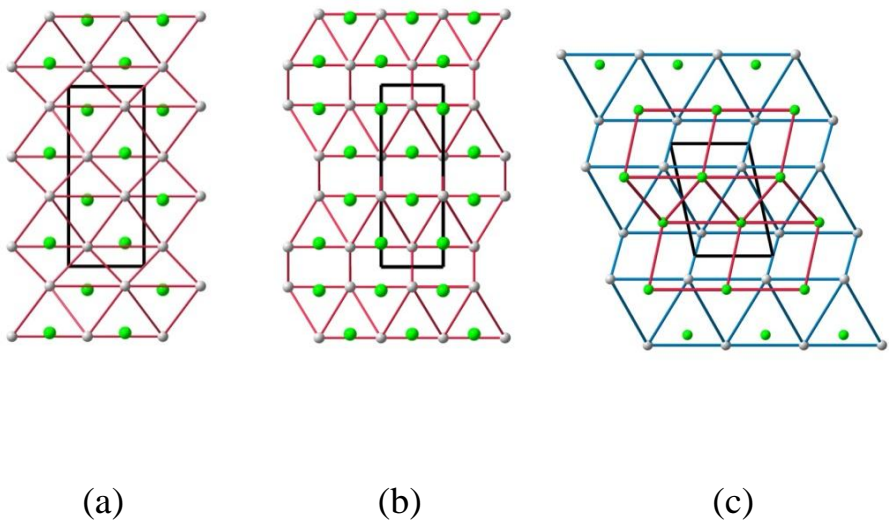

     (a)            (b)            (c)

**Figure 12**

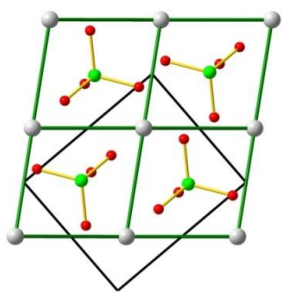 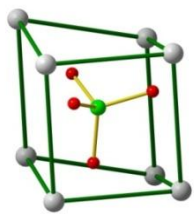 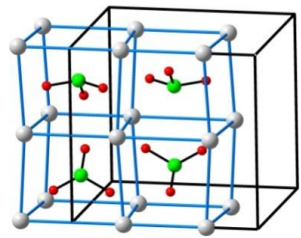

(a) (b) (c)